\documentclass[%
 reprint,
superscriptaddress,
 amsmath,amssymb,
 aps,
]{revtex4-2}

\usepackage{graphicx}
\usepackage{dcolumn}
\usepackage{bm}

\begin{document}

\preprint{APS/123-QED}

\title{Search for Quintessence-Like Pseudoscalar Dark Energy Effects on $^{56}\text{Fe}$ Nuclear Transition Energies in Supernova 1991T}

\author{Robert D. Orlando}
\email{rorlando@irsc.edu}
\affiliation{Department of Physical Sciences, Indian River State College, Fort Pierce, FL 34981, USA}

\author{David S. Koltick}
\affiliation{Department of Physics  and Astronomy, Purdue University, West Lafayette, IN 47907, USA}

\author{Dennis E. Krause}
\affiliation{Physics Department, Wabash College, Crawfordsville, IN 47933, USA}

\date{\today}

\begin{abstract}
The nature of dark energy remains one of the most important unanswered problems in physics. Here we use observations of the Type Ia supernova 1991T to constrain the recent evolution of a dynamical pseudoscalar quintessence-like field $Q(t)$ by comparing the gamma ray spectra emitted by the $^{56}\text{Fe}$ nuclei observed by COMPTEL aboard the Compton Gamma Ray Observatory to terrestrial values. We found that the average fractional energy shift of both the first and second excited states is $\delta E/E = -0.006\pm0.008$, including statistical and systematic errors, indicating that the energies of these astrophysical gamma rays are consistent with the gamma rays produced in terrestrial labs. Assuming that any energy shift is caused by a dynamical QCD axion-like pseudoscalar field $Q(t)$, the observed energy deviations are consistent with a fractional rate of change of the pion mass  at the $68\%$ Gaussian limit given by $ 3\times10^{-11} \geq \dot{m_{\pi}}/m_{\pi}\geq-15\times10^{-11}\text{ yr}^{-1}$. The observed energy deviation was also used to determine the rate of change of the quintessence-like field ($\dot{Q}_0$) for tracking models: $ -1\times10^7\leq \dot{Q}_{\rm 0} \leq 7\times10^7 \text{ GeV/yr}$. This upper limit on $\dot{Q}_0$ results in a fractional kinetic energy $\Omega_{KE}\leq0.029$ and equation of state $-1.0\leq w_{Q}\leq-0.92$, which are consistent with the cosmological constant ($\dot{Q}_0 =0$, $\Omega_{KE}=0$, and $w=-1.0$). Finally, these results were complemented by cosmological observations to place limits on the tracker model decay constant coefficient and tracking power potential.
\end{abstract}

\maketitle


\section{Introduction}

It was first suggested in the 1930s that the universe is expanding after Edwin Hubble observed that the light reaching us from distant galaxies is redshifted \cite{Edwin}. A more surprising discovery came in 1998 when observing the redshift of Type Ia supernovae. Not only is the universe expanding, but the rate of this expansion is increasing \cite{SimpleDEReview}. Later, this fact would be more firmly established by measurements of the power fluctuations in the cosmic microwave background (CMB) \cite{CMBDE}, the comparison of the baryon acoustic oscillations (BAO) signal to theoretical predictions \cite{BAODE}, and the extent to which light is bent by gravitational lensing \cite{GLDE}.

The origin of this acceleration is still a puzzle, but two main paradigms exist to attempt to explain this accelerated expansion. The first is that the force of gravity contains deviations that become measurable at cosmological distance scales  of megaparsecs. Within this paradigm, there is an effort to find a modified theory of general relativity that quantifies and includes these long-range deviations and provides a mechanism by which these deviations are suppressed at smaller distances of astronomical units. Some examples of these extensions are $f$(R) models \cite{FRIntro}, the modified Newtonian dynamics (MOND) theory \cite{MONDIntro}, and the Dvali-Gabadadze (DGP) model \cite{DGPIntro}. 

The second paradigm posits that there is an unaccounted-for energy density, referred to as dark energy, that drives the expansion of the universe \cite{SimpleDEReview}.  The source and nature of dark energy are not known, but cosmological observations suggest that it comprises about 70\% of the energy density of the universe and gives rise to a negative pressure (which corresponds to an equation of state parameter $w_{\Lambda} = -1$) \cite{SimpleDEReview}.

The first and simplest attempt to model dark energy was the reintroduction of the cosmological constant to Einstein's equations\cite{PhenoDE,SimpleDEReview,DEModels}. This led to the cosmological model of Lambda-Cold Dark Matter ($\Lambda$CDM), which is the accepted model of dark energy and dark matter. Within this model, the universe is mostly comprised of a constant vacuum energy that exerts a negative pressure ($-\Lambda$) and drives the accelerated expansion of the universe \cite{Lambda}, and non-relativistic, weakly interacting, pressure-less matter that holds galaxies together (CDM) \cite{CDM}. 

$\Lambda$CDM shows excellent agreement with observation and is considered the most successful cosmological model in physics. However, there are well-documented observational anomalies that are left unexplained as well as theoretical problems that go unsolved with dark energy as a cosmological constant (see \cite{ProbsOfLCDM} for a review). Included in these unaccounted-for anomalies are the Hubble Tension \cite{HubbleConstTensInPersp} and cosmic birefringence \cite{NewBirefringence}. Examples of the unsolved theoretical problems are the cosmological constant problem (the 120 order of magnitude disagreement between theoretical and observed values), the coincidence problem, and the naturalness problem \cite{CCAnthropicPrinciple}.

To attempt to resolve the observational anomalies and solve the theoretical problems of the cosmological constant, dynamical models of dark energy have been introduced. The simplest of these replaces the cosmological constant by a dynamical scalar field with the stipulation that it behaves similarly to a cosmological constant at the current epoch \cite{Wetterich,QuintAMiniRev,QuintARev}. With this model, referred to as quintessence, dark energy is able to change with the evolution of the universe, which introduces a time-dependent equation of state \cite{QuintARev}. Furthermore, depending on the form of the potential of this field, tracking behavior can be exhibited by dark energy where the field's energy density evolves alongside and in relation to the energy densities of matter and radiation \cite{TrackerReview}. 

Most models of quintessence involve scalar fields, but evidence that the cosmic microwave background may be birefringent \cite{NewBirefringence} suggests that a quintessence model based on pseudoscalar fields could provide a natural explanation.  Light and ultralight pseudoscalars are good candidates for the source of dark matter \cite{AxionDarkMatterScience,marsh2024dark,jackson2023search} as well as dark energy \cite{girmohanta_PRD,luu_2025dynamicaldarkenergyultralight,nakagawa2025interpretingcosmicbirefringencedesi}.   Assuming that this new dynamical degree of freedom acting as dark energy is a pseudoscalar and couples to the known fields of the universe introduces the possibility that the fundamental constants of physics can vary over cosmic evolution \cite{VarOfConstants}. This possibility has been explored for the case of dark matter, introducing a variation to the pion mass which would be observed in atomic transition frequencies \cite{ULDMEffectOnAtomicSpectra}. 

A similar scenario is explored here in which an extremely light pseudoscalar dark energy field couples to the hadronic sector, resulting in the change of the nuclear charge radius through its contribution to the pion mass. A probe local to Earth can be used to indicate the temporal variation in this field. For this purpose, we have used the results of the Oklo natural nuclear reactor which uses the disappearance of $^{149}_{62}$Sm to place a limit on the pion mass variation \cite{FlambaumOklo}. Additionally, we use an astrophysical probe to indicate the spatial variation of the dark energy field. In the astrophysical case the change in the charge radius can, in principle, be observed through a deviation in astrophysical gamma-ray energies emitted by nuclear deexcitation when compared to the same deexcitation energies that are observed terrestrially. This energy deviation would be a combination of the temporal and spatial variation in the dark energy field. The astrophysical probes chosen for this purpose are Type Ia supernovae \cite{IaBackground}.

Prompted by the anomalies and problems discussed above, we assume that dark energy exists in the form of dynamical pseudoscalar field with the inverse power-law tracking potential of Ratra and Peebles. We use the possible contribution of this pseudoscalar field to the QCD vacuum angle to, ultimately, derive an upper-limit on the present variation of the dark energy field ($\dot{Q}_0$). Also derived is an upper-limit on the pion mass variation due to its possible applicability across different experiments. Quantities are derived using two sets of measurements: (i) the Oklo results using neutron capture on Samarium and (ii) the spectroscopic gamma-ray data of the peculiar Type Ia SN 1991T which is at a distance of 44.0 million light years from Earth. This is the best supernova for this study because of the availability of its statistically significant data and its distance from earth. To determine if the possible values of $\dot{Q}_0$ are viable and consistent with cosmological observations, we derive the kinetic energy density ($\Omega_k\equiv\dot{Q}_0^2/2\rho_C$) and equation of state ($w_Q\equiv p_Q/\rho_Q$) to generate an allowed parameter space for the free parameters of the field ($\alpha$, the decay constant coefficient and $p$, the power of the Ratra-Peebles tracking potential).

The use of an extra-galactic probe, namely SN 1991T, enables the spatial and temporal variation of the dark energy field to contribute to the gamma-ray energy deviation and the pion mass variation. This is in direct contrast to other experiments that produce a limit on the pion mass variation using an Earth-bound, local probe such as the Oklo natural nuclear reactor where only the temporal evolution of the field can be determined \cite{FlambaumOklo}. The spatial variation of the dark energy field would be caused by inhomogeneities in the distribution of dark energy throughout the universe \cite{InhomoDE, DEInhomo}. This is a scenario that must be taken into account since SN 1991T is at a distance of 13.5 Mpc and the isotropy and homogeneity of the universe only becomes apparent at distance scales of $\sim 100$ Mpc \cite{Ryden}. 

The rest of this paper is structured as follows. We begin with a general discussion of pseudoscalar quintessence models involving a quintessence field $Q(t)$ before focusing on a general form of the tracking potential due to Ratra and Peebles. Since our ultimate goal is to constrain the evolution of the quintessence-like field ($\dot{Q}_0$), we will then simplify the model by Taylor expanding $Q(t)$ over times much shorter than the age of the universe.  Then we investigate the effect of $Q(t)$ on the pion mass, which, in turn, affects nuclear energy levels.  To probe these effects, we turn our attention to Type Ia supernovae, in particular SN 1991T.  Using several nuclear models, we estimate the effects of $Q(t)$ on the gamma-ray spectra emitted by SN1991T, and use the observed spectra by COMPTEL to set constraints on the evolution of the field and the free parameters associated with the evolution. We then conclude with a discussion of the implications of the results.

\section{Quintessence}

The simplest theoretical generalization to the cosmological constant to address its theoretical problems and the unresolved observational anomalies is a quintessence model \cite{DEInPractice,QuintInfRev}. Quintessence is a model where the dynamics of a cosmic scalar field with a minimal coupling to gravity is responsible for the accelerating expansion of our universe \cite{PhenoDE,DEInPractice,DEModels}.

\subsection{General Dynamical Equations}

In Quintessence, the total action with non-relativistic matter is
\begin{equation}
    S = \int d^4x\sqrt{-g}\left[\frac{1}{2}\frac{M_{\rm pl}^2}{8\pi}R-\frac{1}{2}g^{\mu\nu}\partial_{\mu}Q\partial_{\nu}Q-V(Q)\right] + S_{m},
\end{equation}
where $Q$ is the quintessence field, $M_{\rm pl}=1/\sqrt{G}$ is the  Planck mass, $g$ is the determinant of the metric tensor $g_{\mu\nu}$, $R$ is the Ricci scalar, and $S_m$ is the action associated with matter \cite{QuintInfRev,QuintARev}. Throughout the paper, we assume $\hbar = c = 1.$

The resulting equation of motion in the flat, FLRW universe is
\begin{equation}
    \Ddot{Q} + 3H\dot{Q} + \frac{dV}{dQ} = 0,
\label{eq:QuintEoM}
\end{equation}
where $H\equiv\dot{a}/a$ \cite{DEInPractice, DEModels, DEGreatReview}. All quantities with a dot are derivatives of that quantity with respect to cosmic time.

The energy-momentum tensor can be found from the action and can be used to find the energy density and pressure of the field for the quintessence equation of state parameter $w_{Q}$ \cite{DEInPractice}:
\begin{equation}\label{eq:QEdensity}
    \rho_{Q} = -T_0^{0(Q)} = \frac{1}{2}\dot{Q}^2 + V(Q),
\end{equation}
\begin{equation}
    P_{Q} = \frac{1}{3}T_i^{i(Q)} = \frac{1}{2}\dot{Q}^2 - V(Q),
\end{equation}
\begin{equation}\label{eq:wQ}
    w_{Q} \equiv \frac{P_{Q}}{\rho_{Q}} = \frac{\dot{Q}^2/2 - V(Q)}{\dot{Q}^2/2 + V(Q)}.
\end{equation}
It can be surmised from the above expression that the value of $w_{Q}$ corresponds to the cosmological constant case, $w_{\Lambda} \equiv -1$, when $\dot{Q} = 0$ and only the potentials remain in the numerator and denominator. 

Even when studying quintessence models, it is necessary to require that $-1.0 \leq w_{Q} \lesssim -0.972$ at the current time in cosmic history due to the amount of observational evidence that suggests this scenario \cite{Seo,Brout_2022,descollaboration2024darkenergysurveycosmology,rubin2024unionunitycosmology2000,PlanckCollab}. This implies that the ``kinetic energy" of the quintessence field satisfies $\dot{Q}^2/2 \ll V(Q)$, and so is ignored relative to the potential $V(Q)$ when determining the term of the field that corresponds to the cosmological constant (referred to as $Q_{\Lambda}$ later).

Now, we, like others, depart from the canonical quintessence model by assuming that the dynamical field $Q(t)$ is a pseudoscalar field as opposed to a scalar field. In this case, the observation of cosmic birefringence, which is the rotation of the polarizations of CMB photons, can be explained by the existence of a cosmic pseudoscalar field with a Chern-Simons coupling to photons of the form that arises naturally in models of axion-like particles (ALPs) \cite{CosmBiModDep}:
\begin{equation}
    \mathcal{L}_{CB} = -\frac{1}{4}g_{CB}Q F_{\mu\nu} \Tilde{F}^{\mu\nu}.
\end{equation}
Here $F_{\mu\nu}$ is the electromagnetic tensor, $\Tilde{F}_{\mu\nu}$ is its dual, $g_{CB}$ is a coupling constant, and $Q$ is the pseudoscalar field. Furthermore, in order for the non-vanishing vacuum energy to produce the observed cosmic acceleration, the effective mass of the field boson must satisfy \cite{AlphaDisc}
\begin{equation}
    m_Q \leq H_0 \sim 10^{-33} \text{ eV}.
    \label{m_Q mass}
\end{equation}
A field with these characteristics has been shown, using string theory, to be consistent with current cosmological observations \cite{AxionsAsQuintessence}. For these reasons, we will assume that dark energy arises from an ALP pseudoscalar field, with quanta having mass satisfying Eq.~(\ref{m_Q mass}).

\subsection{Choice of Tracking Potential}

In order to preserve the behavior of the quintessence field that can naturally resolve the coincidence and naturalness problems of the cosmological constant, we utilize tracker potentials. The original and simplest tracker potential derived by Ratra and Peebles is an inverse power law that has the general form \cite{OriginalTracker,QuintAMiniRev} 
\begin{equation}\label{eq:TrackerPotential}
    V_{\rm RP}(Q) = M_{\rm RP}^{4+p}Q^{-p},
\end{equation}
where $M_{\rm RP}$ is an energy scale that is fixed by the observed fractional energy density of dark energy $\Omega_{\Lambda} = 0.685$ \cite{DEModels,AstroConsParams}. The main feature of a potential of this form is the quintessence field \textit{tracks} the evolution of the matter and radiation energy densities, which, in turn, influences the evolution of the energy density of the quintessence field. In fact, it is found that the energy density of the quintessence field, with the potential of Eq.~(\ref{eq:TrackerPotential}) can decrease less rapidly than the matter and radiation energy densities of the universe \cite{QuintAMiniRev}. This leads to a point in cosmic history when the energy density of the quintessence field overtakes the energy densities of matter and radiation and becomes the main contributor to accelerated universal expansion, providing a natural resolution to the coincidence problem. Another attractive feature of tracker models is that they are insensitive to initial conditions; a wide range of initial conditions lead to the same cosmic evolution \cite{DEModels,Caldwell}.

\subsection{Simplifying Assumptions}

As previously stated, we are analyzing nuclear capture reactions taken 1.8 billion years ago locally and gamma-ray spectra from SN 1991T, which is 44.0 million light-years away in another galaxy. Therefore, for our purposes, it is sufficient to Taylor expand the quintessence field $Q(t)$ about the current cosmic time $t_{0}$
\begin{equation}\label{eq:TaylorExp1}
    Q(t) = Q(t_{0}) + \dot{Q}(t_0)(t-t_0) + \cdots,
\end{equation}
since $t_{0}- t \ll 1/H_{0}$, where $t_{0} - t$ is of order $10^{7}$~yr and $1/H_{0}$ is of order  $10^{10}$~yr.  For the remainder of this paper, we will assume
\begin{equation}\label{eq:TaylorExp}
    Q(t) \simeq Q_{\Lambda} + \dot{Q}_{0}(t-t_0),
\end{equation}
where the zeroth order term, $Q(t_{0}) = Q_{\Lambda}$, corresponds to the cosmological constant, and 
\begin{equation}
    \dot{Q}_{0} \equiv \dot{Q}(t_{0}) = \left.\frac{dQ}{dt}\right|_{t = t_{0}}
\end{equation}
is the present value of the time derivative of the field with respect to cosmic time. This derivative includes both the spatial and temporal variation of the field since natural units are utilized. The cosmological constant serves as a good approximation of dark energy at the current epoch, which implies that $\dot{Q}$ varies slowly in time and space and can be seen as approximately constant over the time interval investigated here. 

The simple and generic Taylor series expansion given by Eq.~(\ref{eq:TaylorExp}) suggests three non-trivial facts: (i) $\dot{Q}_{0}$ can be seen as a measure of how much dark energy deviates from the cosmological constant in our current epoch, (ii) the larger the value of $t_0-t$, the more stringent limits can be placed on $\dot{Q}_{0}$, (iii) any limits placed on $\dot{Q}_{0}$ can help discriminate among different quintessence models given by Eq. (\ref{eq:QuintEoM}) (iv) the use of an astrophysical probe includes the spatial variation of the field into $\dot{Q}_{0}$ (since $c=1$) .  Of course, for point (ii), $t_{0}-t$ cannot be too large unless higher-order terms are included in the expansion. This expansion permits the derivation of an upper-limit on $\dot{Q}_0$ using recent (i.e. 44.0 million years) cosmic events, but also the use of the Oklo results (1.8 billion years) which is still a small fraction of the universe's age. Both measurements exclude the possibility of discriminating among quintessence-like models since the behavior of the field at the current epoch is insensitive to the initial conditions imposed on the field. It is hoped that more extreme astrophysical events (e.g., additional Type Ia supernovae) than considered in this paper can be utilized to sharpen limits and exclude models.

\subsection{Coupling}

In this paper, we assume that $Q(t)$ is a QCD-like axion field such that it couples to gluons via the QCD vacuum term \cite{ULDMEffectOnAtomicSpectra}
\begin{equation}\label{eq:QCDcoupling}
    \mathcal{L}_{\theta} = \theta\frac{g_S^2 }{32\pi^2} \Tilde{G}^{l\mu\nu} G_{l\mu\nu},
\end{equation}
where $g_S$ is the coupling constant of the strong force, $\theta$ is the QCD vacuum angle, $G_{l\mu\nu}$ is the gluon field strength tensor, and $\Tilde{G}^{l\mu\nu}$ is its dual. We see that $Q(t)$ is not explicit in the equation above; however, just like the QCD axion, a new light pseudoscalar (ALP) field can, in principle, contribute to the small value of $\theta$ \cite{AxionPlusALP}. The mass of the pion, which characterizes the interaction of nucleons via the strong force through virtual pion exchange, can be derived from the larger QCD Lagrangian and results in \cite{PionMassRelationToVacuumAngle}
\begin{equation}\label{eq:PionMass}
    m_{\pi}^2 = \frac{2B_0}{f_{\pi}^2}(m_u^2+m_d^2+2m_um_d\cos\theta)^{1/2}.
\end{equation}
Here, $B_0$ is a constant determined by the ratio of meson masses and has value $7.6\times10^6 \text{ MeV}^3$, $m_u$ and $m_d$ are the masses of the up and down quarks taken to be 4 MeV and 7 MeV, respectively, and $f_{\pi}$ is the pion decay constant, which has value 92.4 MeV. Small QCD vacuum angles not equal to zero result in a pion mass deviation of \cite{ULDMEffectOnAtomicSpectra}
\begin{equation}\label{eq:PionMassVariation}
    \frac{\delta m_{\pi}}{m_{\pi}} \approx -0.05 \theta^2.
\end{equation}
Attributing the entire value of the QCD vacuum angle $\theta$ to $Q$,
\begin{equation}\label{eq:QCDandQ}
    \theta = \frac{Q}{f_{Q}},
\end{equation}
where $f_{Q}$ is the axion decay constant, implies
\begin{equation}
    \frac{\dot{m}_{\pi}}{m_{\pi}} \approx -\frac{0.05}{f_{Q}^{2}}\frac{dQ^{2}}{dt}  \approx -0.1\frac{Q_\Lambda \dot{Q}_{0}}{f_Q^2}.
\end{equation}
For the last term, we have used Eq.~(\ref{eq:TaylorExp}) and kept only the leading order term. The dependence of the pion mass on the time-evolving part of the quintessence-like field is now explicit.

Considering the pion mass determines nuclear properties, including the nuclear mass and radius \cite{ULDMEffectOnAtomicSpectra}, the variation of the pion mass by the quintessence-like field would induce a variation in the nuclear radius through the relation \cite{QuarkMassVar}
\begin{equation}\label{eq:RadiusPion}
    \frac{\delta r_0}{r_0} = 1.2\frac{\delta m_{\pi}}{m_{\pi}},
\end{equation}
where $r_0 = 1.2$~fm. The above equation takes into account the strength of the strong interaction, indicated by the mass of the pion, and how readily the constituent nucleons respond to that force, indicated by the mass of the individual nucleons. The resulting change in the nuclear radius over time can now be related to the quintessence-like field and yields the expression
\begin{equation}
    \frac{\dot{r}_0}{r_0}\equiv\frac{1}{\Delta t}\frac{\delta r_0}{r_0} = -0.12\frac{Q_\Lambda \dot{Q}_{0}}{f_Q^2}.
\end{equation}
This equation can now be used to estimate the effects of $\dot{Q}_{0}$ on nuclear energy levels and gamma-ray spectra.

\section{Astrophysical Probe and Considerations}

The fact that the quantity $\dot{Q}_{0}$ is small relative to $Q_\Lambda$ in the present epoch indicates that there are three approaches to measuring it: (i) high-precision terrestrial experiments, (ii) long-term terrestrial experiments, (iii) and finally, our motivation, the investigation of gamma-ray spectra from extragalactic energetic astrophysical systems that can be compared with present day terrestrial measurements. Any differences in these spectra might be attributed to a spatial or temporal change, or both, in the dark energy field. The COMPTEL data collected viewing SN 1991T will be used to study the spatial change in dark energy and to complement the temporal change in dark energy indicated by Oklo.

\subsection{Nuclear Processes Within Type Ia Supernovae Producing Their Gamma-Ray Spectra}

The detonation of a white dwarf producing a standard or peculiar Type Ia supernova creates an extremely energetic environment in which a population of radioactive nickel $^{56}\text{Ni}$ forms. This population of $^{56}\text{Ni}$ will decay into $^{56}\text{Co}$ by electron capture, which, in turn, decays into $^{56}\text{Fe}$, also via electron capture \cite{IaBackground}:
\begin{equation}
    ^{56}\text{Ni} \rightarrow ^{56}\text{Co} \rightarrow ^{56}\text{Fe}.
\end{equation}
Soon after the explosion, the resulting ejecta is optically thick to the gamma rays produced by these nuclear decays, resulting in gamma-rays that are down-scattered to UV, optical, and IR energies. Only 10--30 days after the explosion, will the ejecta be optically thin and the gamma-rays be directly observable \cite{SN2014J}. 

Since $^{56}\text{Ni}$ has a half-life of 6 days (see \cite{NuDat}), the gamma rays produced by these decays cannot be directly observed. Rather, a gamma-ray spectrum will only be available once the population of $^{56}\text{Co}$, which has a half-life of approximately 77 days (see \cite{NuDat}), is decaying into $^{56}\text{Fe}$. The decay of $^{56}\text{Co}$ into $^{56}\text{Fe}$ occurs with an intensity of 100\% and produces $^{56}\text{Fe}$ in an excited state 100\% of the time \cite{NuDat}. 

Of interest in this work is the $^{56}\text{Fe}$ gamma-ray de-excitation spectra dominated by the 1238 keV line, the transition from the second excited 4+ state to
the first excited 2+ state with branching ratio 66.46\% and by the 847 keV line
from the transition of the first excited 2+ state to the 0+ ground state, with
branching ratio 99.94\% \cite{NuDat,IronDeexcitation}. There also exists a transition that produces a gamma-ray of 1038 keV with a relative intensity of 14\%, which was included in our best fits.

\subsection{SN 1991T}

The astrophysical event utilized for this study is SN 1991T, a peculiar Type Ia supernova detected on April 13, 1991, in the host galaxy NGC 4527 observed by COMPTEL \cite{91TBackground}. Data was collected for this supernova for two 14-day periods 66 and 176 days after the explosion. SN 1991T was chosen on account of the availability of appropriate gamma-ray spectroscopy and the fact that its distance from Earth (44.0 million light-years) was significant, but not too large so that $t_{0} - t \ll 1/H_{0}$.

Type Ia supernovae are characterized by the absence of hydrogen and helium in their spectra with a distinct absorption line near 6100~\AA\ \cite{IaBackground}. However, due to the peculiar properties of SN 1991T, a new subtype of Type Ia's referred to as 91T-like supernovae was introduced. Supernovae of this subtype (along with 91T) have higher luminosities and less homogeneity in their light curves and spectra compared to those of standard Type Ia's \cite{91TPeculiar}.  

In general, the detonation of a white dwarf producing a Type Ia supernova, regardless of any initial asymmetry, is expected to create a spherically symmetric distribution of $^{56}\text{Ni}$ after the ejecta is optically thin \cite{SymmetricNi56Layer}. Since SN 1991T was observed with an above average luminosity, it was explored if this observed charactersitic would be caused by an asymmetric ejecta shape. It was concluded that SN 1991T is a marginal case of a spherical explosion \cite{91TSpeedDecay}. 

The combined spectral data of SN 1991T implies the decay of $^{56}$Co with a significance above $3\sigma$ based on the observed $^{56}$Fe lines \cite{91TData}. However, this claim has been contested in the literature (see \cite{91TContested} and the references therein). The disagreement arises from the observed $^{56}$Fe gamma-ray fluxes being too small relative to the mass of $^{56}\text{Ni}$ that SN 1991T is believed to contain based on its luminosity \cite{Contention1,Contenion2}. The smaller flux cannot be predicted with any known model of Type Ia supernovae. In contrast to this objection, it should be noted that Reference \cite{91TData} was produced after the publication of References \cite{Contention1} and \cite{Contenion2} to address any systematic errors that would produce a signal that mimics the decay of $^{56}$Co. It was concluded by the authors that the 3.3$\sigma$ signal does not overstate the confidence that there are $^{56}$Co decays occurring in SN 1991T, suggesting that the objections have been resolved.

\subsection{COMPTEL}

The instrument that observed SN 1991T was the imaging Compton telescope COMPTEL, one of four instruments aboard the Compton Gamma Ray observatory. The observatory orbits the earth at a radius of 450 kilometers. COMPTEL explores phenomena within the energy range of $\sim$1--30 MeV by recording coincidences between an upper detector array ($D_1$) constructed with NE213 liquid scintillator and a lower NaI (TI) detector array ($D_2$) \cite{COMPTELCatalogue}. The quantities measured are the energy deposited in $D_1$ and $D_2$, the time of flight between $D_1$ and $D_2$, the location of the interaction in the two detector arrays, the pulse shape in $D_1$, and the time of the entire event. The final two quantities in this list are used for background discrimination \cite{COMPTELDescription,COMPTELCatalogue}. The instrument is calibrated using the onboard decay of $^{60}\text{Co}$, producing gamma rays of 1.17 MeV and 1.33 MeV \cite{COMPTELDescription}. For a full description of this process, see reference \cite{SNELLINGCalibration}.

\subsection{Nuclear $^{56}\text{Fe}$ Line Shape}

The gamma-ray lines that correspond to Dirac peaks in the laboratory will appear as curves with widths based on the velocity of the ejecta. Once the ejecta is optically thin, assuming that the radioactive layer is uniform and expanding isotropically, the line shape is expected to be a symmetric parabolic curve. It is known that the ejecta of SN 1991T initially expands with a speed of $\sim10,000$ km/s \cite{91TSpeedDecay}. The rms variance of the parabolic line shape is estimated to be
\begin{equation}
    \sigma_v = \frac{20,000 \text { km/s}}{\sqrt{20}}\approx4470 \text{ km/s} ,
\end{equation}
and results in an energy variance ratio
\begin{equation}
    \frac{\sigma_e}{E}=\frac{1}{2}\left(\frac{1}{1-z} -\frac{1}{1+z}\right) \approx0.015.
\end{equation}

To find the observed line shape, the energy resolution of the COMPTEL detector system must be convoluted with a parabolic line shape. To do so, we compare to the width of the Gaussian COMPTEL $^{56}\text{Fe}$ line energy resolution, found to be
\begin{equation}
    \frac{\sigma_c}{E}\approx0.04.
\end{equation}
Given the data's statistics and the detector's energy resolution, 47 keV at the position of the $^{56}\text{Fe}$(1238) line, the natural parabolic line shape of 6 keV is not resolvable by COMPTEL and is ignored. Therefore, the data is fit using the detector's resolution function with a Gaussian line shape in agreement with other analyses of the COMPTEL data \cite{91TData}.

\section{Nuclear Models}

Type Ia supernovae contain the decay of $^{56}$Co into an excited state of $^{56}$Fe by the process of electron capture. Two models are used and compared to calculate the effect of the quintessence field on the nuclear levels. The first, assumes these excited states of $^{56}$Fe, an even-even nucleus, are described as collective nucleon motion modeled as a surface phonon on a charged liquid drop. The second, also a collective model, assumes the excited nucleus is a deformed rigid rotor. The two models are compared to estimate the theoretical systematic uncertainty in $\dot{Q}_0$.

\subsection{Liquid Drop Model}

The low-lying excited states of even-even nuclei are understood as surface excitations of a nuclear fluid \cite{Erba}. The excited state energies of interest are characterized by the number of phonons $N$ with frequencies $\omega_{\lambda}$. The surface of the nucleus is given in terms of the spherical harmonic functions with time-dependent deformation parameters \cite{Preston}. Assuming that the deformation parameters are small, the Hamiltonian for this system takes on the form of a harmonic oscillator \cite{Erba,Preston}. 

The coefficients of this Hamiltonian are what determine the frequency of the vibrational deformation \cite{Preston}. The frequency of the phonons is given by
\begin{equation}
    \omega_{\lambda} = \sqrt{\frac{\|C_{\lambda}\|}{B_{\lambda}}},
\end{equation}
where $B_{\lambda}$ is the inertia of the nucleus and $C_{\lambda}$ is the restoring force of the potential. We choose $\lambda=2$ which is the smallest allowed value for $\lambda$ and corresponds to quadrupole variations of the nuclear surface. The values of $\lambda=0,1$ are excluded since the nucleus is assumed to be incompressible and translations of the center of mass do not correspond to surface deformations [62].

For an irrotational incompressible fluid these coefficients were found by Rayleigh to be \cite{Preston,Rayleigh}
\begin{equation}
    B_{\rm \lambda,irr} = \frac{\rho R_0^5}{\lambda}, \quad C_{\rm \lambda,irr} = C^{(1)}_{\lambda} - C^{(2)}_{\lambda},
\end{equation}
where 
\begin{eqnarray}
    C^{(1)}_{\lambda} & = & SR_0^2(\lambda-1)(\lambda+2),
    \\
    C^{(2)}_{\lambda} & =& \frac{3}{4\pi}\frac{Z^2e^2}{R_0}\frac{\lambda-1}{2\lambda+1}.
\end{eqnarray}
Here, $\rho$ is the nuclear mass density for the spherical configuration, \textit{e} is the fundamental charge, and $S$ is the surface energy per unit area \cite{Feshbach}. The value of $S$ is found using the coefficient of the $A^{2/3}$ term in the semi-empirical mass formula. 

Using the known values of the mass density and only considering quadrupole vibrational deformations ($\lambda=2$), the coefficients become
\begin{equation}\label{eq:PhononCoefficients}
    B_{\rm 2,irr} = \frac{3}{8\pi}m_{p}AR_0^2, \quad C_{\rm 2,irr} = 4R_0^2S-\frac{3}{20\pi}\frac{Z^2e^2}{R_0}.
\end{equation}
The quantization of these vibrations leads to phonons, which determine the energies of the excited states through \cite{Preston}
\begin{equation}\label{eq:PredictedEphon}
    E_N = N\hbar\omega_2,
\end{equation}
where $N = 0, 1, 2, \ldots$ is the number of phonons \cite{Erba}. Relevant to this study are the first and second excited states of $^{56}$Fe, which coincide with $N=1$ and $N=2$ respectively. In Appendix  \ref{app:PhonDeriv}, we show the phonon model yields the fractional time rate of change of the nuclear energy levels during the time interval $\Delta t = t -t_{0}$ due to $Q(t)$ is given by
\begin{eqnarray}
    \label{eq:EDeviationPhon}
    \frac{ \dot{E}}{E} & \equiv &  \frac{1}{\Delta t}\frac{\delta E}{E} 
    \nonumber \\
    & = & -\frac{(1-3\xi/2)}{(1-\xi)}\frac{\dot{r}_0}{r_0}
    \nonumber \\
    & = & 0.12\frac{(1-3\xi/2)}{(1-\xi)}\frac{Q_\Lambda\dot{Q}_{0}}{f_Q^2},
\end{eqnarray}
where
\begin{equation}
    \xi \equiv \frac{3}{10} \frac{Z^2e^2}{\varepsilon A}\frac{1}{r_0}.
\end{equation}
Here, $\varepsilon \approx  18.56$ MeV is the nuclear surface energy, which is determined by the semi-empirical mass formula \cite{Feshbach}.  For the $^{56}\text{Fe}$ nucleus, one finds $\xi = 2.94$.

Finally, Eq.~(\ref{eq:EDeviationPhon}) can be combined with Eq.~(\ref{eq:RadiusPion}) to determine the fractional rate of change of the pion mass due to $Q(t)$ assuming the liquid drop model:
\begin{equation}\label{eq:PionDeviationPhon}
 \frac{\dot{m}_{\pi}}{m_{\pi}} = \frac{1}{1.2}\frac{\dot{r}_{0}}{r_{0}} \approx -\frac{1}{1.2}\frac{(1-\xi)}{(1-3\xi/2)}\frac{1}{\Delta t}\frac{\delta E}{E}. 
\end{equation}

\subsection{Rigid Rotor Model}

The same quantities are presented here, as above, but for the excited nucleus modeled as a deformed rigid rotor. These are determined to be
\begin{equation}\label{eq:EDeviationRig}
    \frac{\dot{E}}{E}\equiv \frac{1}{\Delta t}\frac{\delta E}{E}= -2\frac{\dot{r}_0}{r_0} = \frac{0.24}{f_Q^2}Q_{\Lambda}\dot{Q}_{0}, 
\end{equation}
and
\begin{equation}\label{eq:PionDeviationRig}
 \frac{\dot{m}_{\pi}}{m_{\pi}} = \frac{1}{1.2}\frac{\dot{r}_{0}}{r_{0}} \approx -\frac{1}{2.4\Delta t}\frac{\delta E}{E}. 
\end{equation}
For a short derivation of Eq. (\ref{eq:EDeviationRig}), refer to Appendix \ref{app:RigDeriv}. For more detail on the model and how it applies to the $^{56}$Fe nuclei within SN1991T, see Reference \cite{OrlandoThesis}.

\subsection{Time Variance of the Quintessence-like Field}

The results from the previous subsection provides an avenue for constraining the present-day value of the rate of change of the ALP quintessence-like field, $\dot{Q}_0$, by searching for variations in nuclear gamma spectra $\delta E/E$ over a time interval $\Delta t$.  Re-arranging Eqs. (\ref{eq:EDeviationPhon}) and (\ref{eq:EDeviationRig}), we obtain  $\dot{Q}_{0}$ from the liquid drop and rigid rotor models, respectively:
\begin{equation}\label{eq:Q0Phon}
    \dot{Q}_{\rm 0,phon} = \frac{f_Q^2}{0.12Q_{\Lambda}\Delta t}\frac{(1-\xi)}{(1-3\xi/2)}\frac{\delta E}{E}.
\end{equation}
\begin{equation}\label{eq:Q0Rig}
    \dot{Q}_{\rm 0,rig} = \frac{f_Q^2}{0.24Q_{\Lambda}\Delta t}\frac{\delta E}{E}, 
\end{equation}
 We find that the final results are relatively insensitive to which of these models is used. The ratio of the two quantities yields
\begin{equation}
    \frac{\dot{Q}_{\rm 0,rig}}{\dot{Q}_{\rm 0,phon}} = \frac{1}{2}\frac{(1-3\xi/2)}{1-\xi} \approx 0.88,
\end{equation}
indicating that $\dot{Q}_{\rm 0,phon}$ is larger than $\dot{Q}_{\rm 0,rig}$ by only $12\%$. Therefore, $\dot{Q}_0$ will be calculated solely with the even-even $^{56}\text{Fe}$ nucleus modeled as a liquid drop.

Before explicit values can be derived for $\dot{Q}_0$ for the liquid drop model, the values of $Q_\Lambda$ and $f_Q$ must be determined. A majority of the measurements of the dark energy equation of state parameter are consistent with a cosmological constant ($w_{\Lambda} = -1$) with some possibility for the equation of state to deviate from this value \cite{Brout_2022,rubin2024unionunitycosmology2000,descollaboration2024darkenergysurveycosmology}. To be consistent with this observation, it is assumed that the energy density of the quintessence field, given by Eq. (\ref{eq:QEdensity}), is dominated by the quintessence potential $V(Q)$
\begin{equation}\label{eq:ValueOfQLamb}
    \rho_{Q} \approx V(Q_{\Lambda}).
\end{equation}
Using the above equation and the Ratra-Peebles tracking potential of Eq. (\ref{eq:TrackerPotential}) implies that the constant term of the quintessence field is
\begin{equation}\label{eq:ConstTermOfQ}
    Q_{\Lambda} = \frac{1}{\sqrt{8\pi G}}\frac{1}{(3\Omega_{\Lambda})^{1/p}} = \frac{M_{\rm pl}}{\sqrt{8\pi }}\frac{1}{(3\Omega_{\Lambda})^{1/p}},
\end{equation}
where $M_{\rm pl} = 1/\sqrt{G}\approx 1.22\times10^{19}$ GeV is the Planck mass. To derive this expression, the tracking potential parameter $M_{\rm RP}$ was set to
\begin{equation}\label{eq:PotentialRequirement}
    M_{\rm RP}^{4+p} \approx (8\pi G)^{-1-p/2}H^2_0,
\end{equation}
to allow for the slow evolution of the quintessence field~\cite{DEModels}.

The results of string theory suggest that a light pseudoscalar degree of freedom can act as dark energy if the mass of that pseudoscalar is $m_{Q}\lesssim 10^{-33}$ eV \cite{AxionsAsQuintessence}. To ensure that this mass is generated by the quintessence field, the decay constant is taken to be
\begin{equation}\label{eq:decayconst}
    f_Q = \alpha M_{\rm pl}/\sqrt{8\pi},
\end{equation}
where $\alpha \sim 0.1$. In fact, string constructions indicate that the coefficient can be as low as $\alpha \sim 10^{-3}$ for dark energy and dark matter \cite{S200}. This is a common choice for ALP quintessence models (see \cite{QuintAxionDE,Axiverse,EWInstatons,CosmicBiAndEWDE,EWAxion,QuintAxionDE}) due to the fact that this is the energy scale with which new physics is associated. Furthermore, ALP quintessence models have two notable successes associated with them. The first is that they are able to reproduce the surmised energy density of dark energy (specifically in the case of Electroweak Dark energy \cite{EWAxion}) of 
\begin{equation}
    S_I = 10^{-122} M_{\rm pl}^4,
\end{equation}
where $S_I$ is the vacuum energy indicated by the action of the Electroweak Axion. The second success of these models is that they provide an opportunity to resolve the Hubble tension \cite{QuintAxionDE} (see Ref.~\cite{HubbleConstTensInPersp} for a review of the Hubble tension). Since this work operates on the basis that dark energy is an ALP field, and to possibly preserve the advantages of similar models, Eq.~(\ref{eq:decayconst}) will be used for the decay constant of the quintessence-like field.

Using the expressions for $Q_{\Lambda}$ and $f_{Q}$ given by Eqs.~(\ref{eq:ConstTermOfQ}) and (\ref{eq:decayconst}), respectively, we  obtain the final expressions for $\dot{Q}_{0}$ for the liquid drop model, Eq.~ (\ref{eq:Q0Phon}), as 
\begin{equation}\label{eq:Q0PhonFinal}
    \dot{Q}_0\equiv\dot{Q}_{\rm 0,phon} = \frac{1}{0.12}\frac{(1-\xi)}{(1-3\xi/2)}
    \left[\frac{\alpha^{2}M_{\rm pl}}{\sqrt{8\pi}}(3\Omega_{\Lambda})^{1/p}\right]
    \frac{1}{\Delta t}\frac{\delta E}{E}.
\end{equation}
Aside from the factors associated with the liquid drop model, these results involve two dimensionless parameters: $\alpha$, which is proportional to the axion decay constant $f_{Q}$, and $p$, the tracking potential power.

\section{COMPTEL Data Analysis}

To determine the fractional energy deviation $\delta E/E$ of astrophysical gamma rays to their terrestrial counterparts, we fit the published COMPTEL data \cite{91TData} of COMPTEL's observation of SN 1991T using \texttt{Igor} \cite{IGOR}. Subsequently, the value of the fractional energy deviation was used to calculate values of $\dot{Q}_0$ utilizing the liquid drop model of the nucleus. 

\subsection{$^{56}\text{Fe}$ Line Centroids and Widths}\label{sec:Fits}

The energy resolution function of COMPTEL is given by \cite{COMPTELDescription} 
\begin{equation}\label{eq:EnergyResFunc}
    \sigma(E) = 0.01\sqrt{14.61E+2.53E^2},
\end{equation}
where $E$ is the line energy in units of MeV. The widths that are calculated for the $^{56}\text{Fe}$ lines of energies 846.8 keV, 1038 keV, and 1238 keV are 38 keV, 42 keV, and 47 keV, respectively. 

The $^{56}\text{Fe}$ line fits are performed on the background-subtracted data, which has a 50 keV bin width. Before the data is fit, consideration must be given to the possible presence of neutron capture on $^{27}\text{Al}$ in the spacecraft generating a $^{27}\text{Mg}$ background in the spectrum. An analysis for $^{27}\text{Mg}$ was undertaken by Morris \textit{et al.} in Ref. \cite{91TData}, but this analysis did not consider the nearby required $^{56}\text{Fe}$(1038) line in exact proportion to the $^{56}\text{Fe}$(1238) line.

\begin{figure}
    \centering
    \includegraphics[width=1.0\linewidth]{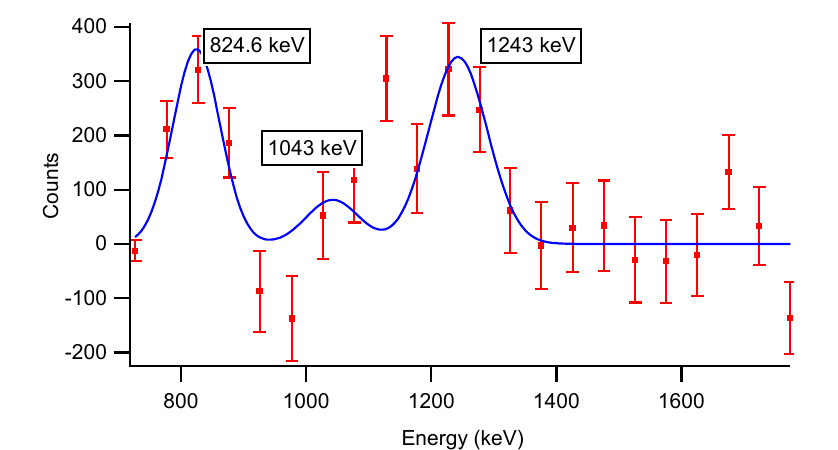}
    \caption{Best fit of the background-subtracted SN 1991T gamma-ray data after fixing Gaussian widths. Negative counts occur when the fitted background counts exceed those of the collected data.}
    \label{fig:91FitDiscussion}
\end{figure}

To check for the presence of the $^{27}\text{Mg}$ line, the spectrum in Fig. \ref{fig:91FitDiscussion} was fit using 5 independent parameters. An independent amplitude and energy for the $^{56}\text{Fe}$(846.8) line and  the $^{56}\text{Fe}$(1238) line, a single amplitude parameter for the $^{27}\text{Mg}$(1014), because its energy is known since the line is local to the spacecraft, and no free parameters for the $^{56}\text{Fe}$(1038) line because its energy and amplitude are exactly known relative to the $^{56}\text{Fe}$(1238) line.

The resulting fit yields a $\chi^2/DoF$ of 2.1 with $0\pm140$ counts for the $^{27}\text{Mg}$(1014) line to be compared to the 1038 keV, 1238 keV $^{56}\text{Fe}$ complex of $962\pm150$ counts. Excluding the $^{27}\text{Mg}$ line for the fit improves the $\chi^2/DoF$ to 1.9. For this reason we do not include the $^{27}\text{Mg}$ line in our analysis.

Fitting the SN 1991T data after fixing the values of the widths to those determined by utilizing Eq. (\ref{eq:EnergyResFunc}) results in the best fit provided in Fig. \ref{fig:91FitDiscussion}. In this fit, the two $^{56}\text{Fe}$ lines are placed  at 824.6 $\pm$ 7.3~keV and 1243 $\pm$ 12~keV, where the uncertainties are one standard deviation. 

The energy shift of the middle iron line is not considered here because it was included without its own independent parameters. The energy of this line was fit to be 200 keV below the upper iron line with 21\% of the upper iron line's counts, which was determined based on this nuclear transition's relative intensity.

\subsection{Doppler Redshift Correction}

The energies of the $^{56}\text{Fe}$ lines are corrected for the Doppler redshift due to the heliocentric radial velocity of NGC 4527. This is determined using the equation
\begin{equation}
    E_{\rm exp} = \frac{E_E}{1+z},
\end{equation}
where $E_{\rm exp}$ is the expected gamma-ray energy, $E_E$ is the observed gamma-ray energy here on earth, and $z$ is the redshift of NGC 4527. The relative velocity of this host galaxy results in a redshift $z$ of 0.005791 \cite{NED}. This shifts the iron lines observed on Earth at 846.8 keV and 1238 keV to 841.9 keV and 1231 keV, respectively \cite{NuDat}.

\subsection{Energy Deviation of $^{56}\text{Fe}$ Lines}

The energy shifts of the $^{56}\text{Fe}$ lines are displayed in Table \ref{tb:FitValues}, along with their Earth and expected energies. In our fit, the upper iron line is shifted upward in energy by 1$\sigma$ while the lower line is shifted downward in energy by $2.3\sigma$. The resulting percentage differences for the first and second excited state energies are
\begin{equation}
    \frac{\delta E_1}{E_1} = -0.020\pm.009, \quad \frac{\delta E_2}{E_2} = 0.010\pm0.010,
\end{equation}
where the uncertainties are $1\sigma$. 

Taking the weighted average in the energy percentage difference and the statistical errors only, the combined energy deviation is
\begin{equation}\label{eq:average delta E/E}
    \frac{\delta E}{E} = -0.007\pm0.007, 
\end{equation}
which is consistent with the expected value of zero.

\subsection{Possible Systematic Shift}

It is important to note the background model that was used to generate the background-subtracted SN 1991T data indicates that there is a 6 keV upward shift of the 2223 keV deuterium line to 2229$\pm$1.18 keV. This line forms due to neutron capture local to the spacecraft \cite{91TData}. This shift may be due to interference with a spectral artifact at energies above 2.5 MeV that is not taken into account in the background model (refer to \cite{91TData} and  \cite{OrlandoThesis}). Despite various attempts to account for the artifact, the upward 6 keV shift remains in all fits of extensions to the background model. 

In the background model, the only calibration line available is the deuterium line. Assuming to first order that the COMPTEL calibration is a linear response function ($E(x)=a+bx$), this 6 keV shift would correspond to  a 6 keV shift in the $^{56}\text{Fe}$ lines if due to an error in the $a$ term calibration. If there is an error in the $b$ term calibration, there is a 2.2 keV shift in the lower iron line and a 3.4 keV shift in the upper iron line. Adding the maximum systematic shift of 6 keV in quadrature with the statistical errors yields the results
\begin{equation}
    E_1 = 824.6\pm9.4\text{ keV},
\end{equation}
\begin{equation}
    E_2 = 1243\pm13\text{ keV}.
\end{equation}
These energies result in the combined energy shift of
\begin{equation}
    \frac{\delta E}{E} = -0.006\pm0.008.
\end{equation}
We will use this value to calculate the pion mass variation, $ \dot{m}_{\pi}/m_{\pi}$ and the time-varying part of the quintessence-like field, $\dot{Q}_{0,max}$.

\begin{table*}
\caption{COMPTEL $^{56}$Fe $\gamma$-ray energies collected from SN 1991T}\label{tb:FitValues}
\begin{ruledtabular}
\begin{tabular}{cccccc}
 &Terrestrial &redshifted &COMPTEL&Difference&Significance \\ \hline
    State of $^{56}$Fe & $E_E$ (keV) & $E_{\rm exp}$ (keV) & $E_{\rm fit}$ (keV) & $\Delta E_{\rm exp,fit}$ (keV) & $\sigma$-value \\ \hline
    Second Excited State & 1238 & 1231 & 1243(13) & 12 & $0.9\sigma$ \\
    First Excited State & 846.8 & 841.9 & 824.6(9.4) & -17.3 & $1.8\sigma$ \\
\end{tabular}
\begin{tabular}{ccc}
          & $\delta E/E$ & Significance \\ \hline
         Combined & -0.006$\pm$0.008 & 0.8$\sigma$\\
    \end{tabular}
\end{ruledtabular}
\end{table*}

\subsection{Time Variation of the Pion Mass}

Using the averaged fractional energy change given by Eq.~(\ref{eq:average delta E/E}) observed from the SN 1991T gamma-ray spectra, we can now investigate its implications for dark energy models.  If we attribute the observed shift entirely due to the change in the pion mass induced by the pseudoscalar quintessence field $Q(t)$, we find for the liquid drop model [Eq.~(\ref{eq:PionDeviationPhon})] 
\begin{equation}
    3\times10^{-11} \geq\frac{\dot{m}_{\pi}}{m_{\pi}} \geq
    -15\times10^{-11} \text{ yr}^{-1},
\end{equation}
where we have used $\Delta t = -44\times10^6$ yrs and  $\xi = 2.94$.

The above limit on the pion mass variation can be compared to the local, time-only quantity as determined by the variation in atomic frequency transitions caused by nuclear properties. One such atom used for this purpose is singly-ionized Ytterbium ($Z=70$), where the drift in E3 and E2 transition frequency ratios
\begin{equation}
    \frac{\delta(\nu_3/\nu_2)}{\nu_3/\nu_2},
\end{equation}
are measured to extract the corresponding pion mass variation (for more details, see \cite{ULDMEffectOnAtomicSpectra} and the references contained therein). In this way, the pion mass variation is found to be \cite{QuarkMassVar}
\begin{equation}
    0.10\times10^{-15} \geq \frac{\dot{m}_{\pi}}{{m}_\pi} \geq -0.52\times10^{-15} \text{ yr}^{-1}.
\end{equation}

Additionally, the limit on the pion mass variation can be derived from analyses of the Oklo natural nuclear reactor. The studies of Oklo are conducted using single particle models and the observed disappearance of $^{149}_{62}$Sm for the goal of generating a limit on the quark mass variation, which results in a pion mass variation of \cite{FlambaumOklo}
\begin{equation}
    \left|\frac{\dot{m}_{\pi}}{{m}_\pi}\right| \leq 1.1\times10^{-18} \text{ yr}^{-1}.
\end{equation}
Note that \cite{FlambaumOklo} gives a result for the quark mass variation from the Oklo analysis. However, the pion mass variation can be surmised from this quantity by multiplying by the sensitivity coefficient of $K_{\pi,q} = 0.498$ \cite{SensitivityCoefficient}. Separate subsequent analyses use these results to indicate the dark matter energy density \cite{QuarkMassVar}.

The limit on the pion mass variation derived in this analysis is five orders of magnitude larger than the atomic clock limit and seven orders of magnitude larger than the Oklo limit. Unlike atomic clock experiments and analyses of Oklo, we have used a nonlocal extragalactic gamma-ray source (SN 1991T) to indicate a pion mass variation. The host galaxy of SN 1991T, NGC 4527, belongs to the M61 group, which is a member of the Virgo II group. This group is located 55 million light years from Earth, putting it well outside our local group \cite{Atlas}. We emphasize here, the non-locality of SN 1991T. The pion mass variation probes the spatial difference caused by inhomogeneities in dark energy \cite{InhomoDE, DEInhomo}, in addition to the temporal effect caused by how the field changes over the evolution of the universe. The atomic clock and Oklo measurements are strictly local, probing only the temporal variation of the field.

\subsection{Limits on \boldmath $\dot{Q}_{0}$}

To obtain the value of $\dot{Q}_0$ for the liquid drop model, we used Eq. (\ref{eq:Q0PhonFinal}) with the values $M_{\rm pl} = 1.22\times10^{19}\text{ GeV}$, $\Delta t = -44\times10^6$ yrs, $\Omega_{\Lambda} = 0.685$, $\xi = 2.94$, and $\delta E/E = -0.006\pm0.008$. We take the upper limit on $\dot{Q}_0$ to be yielded when $p=1$ and $\alpha=0.1$. This is because $p>0$ for the tracking behavior of the Ratra-Peebles potential and $p=1$ most mimics the late-time accelerated expansion of $\Lambda$CDM \cite{GreatTrackerReview}. Additionally, $\alpha \lesssim 0.1$ prevents the kinetic energy density of $Q$ from exceeding 10\% of the total dark energy density. The latter condition is required to be consistent with the approximation made in Eq.~(\ref{eq:ValueOfQLamb}). The value of $\dot{Q}_{0}$ for the liquid drop model gives
\begin{equation}
\dot{Q}_{0}(\alpha,p) \leq   \dot{Q}_{0}(\alpha=0.1,p=1),
\end{equation}
where
\begin{equation}
    \dot{Q}_{0,max}\equiv\dot{Q}_{0}(\alpha=0.1,p=1) = (3\pm4)\times10^7\,\,\mbox{GeV/yr}.
\end{equation}

An alternative constraint more useful for cosmological purposes is found by expressing $\dot{Q}_{0}$ in terms of the kinetic energy density of $Q(t)$.  Defining the fraction
\begin{equation}\label{eq:OmegaKE}
    \Omega_{\rm KE} \equiv \frac{\dot{Q}_{0}^2/2}{\rho_{\rm crit}},
\end{equation}
Eq.~(\ref{eq:wQ}) can be rewritten as
\begin{equation}
    w_{Q} \approx \frac{\Omega_{\rm KE}-\Omega_{\Lambda}}{\Omega_{\rm KE}+\Omega_{\Lambda}},
\end{equation}
where we have used 
\begin{equation}
\Omega_{V} \equiv \frac{V(Q_{0})}{\rho_{\rm crit}} \simeq 
\frac{V(Q_{\Lambda})}{\rho_{\rm crit}} = \Omega_{\Lambda}.
\end{equation}
Solving for $\Omega_{\rm KE}$, we obtain
\begin{equation}\label{eq:OmegaKEQDot}
\Omega_{\rm KE} =\Omega_{\Lambda}\left(\frac{1 + w_{Q}}{1- w_{Q}}\right).
\end{equation}
The above representative value of $\dot{Q}_{0,max}$ using the SN 1991T data yields a maximum fractional kinetic energy density and equation of state of
\begin{equation}
    \Omega_{KE}\leq 0.029, \quad -1.0\leq w_{Q} \leq -0.92,
\end{equation}
which can be compared to the same quantites calculatd using the Oklo pion mass variation
\begin{equation}
    \Omega_{KE,\text{Oklo}}<1.18\times10^{-18}, \quad w_{Q,\text{Oklo}} \approx -1.0.
\end{equation}
The constraint using the SN 1991T data is more permissive than those surmised from other observations.

\begin{figure}
    \centering
    \includegraphics[width=\linewidth]{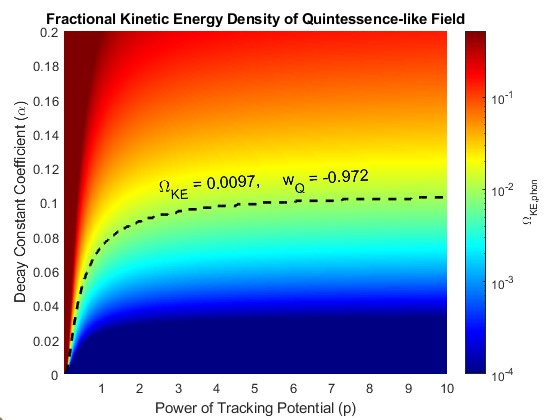}
    \caption{Kinetic energy density $\Omega_{KE}$ for different combinations of the parameters $\alpha$ and $p$. The curve drawn on the figure indicate combinations of $\alpha$ and $p$ that allow for an equation of state parameter $w_{Q} = -0.972$. Below the curve, the value of $w_Q$ decreases to a minimum value of -1.0, which is consistent with dark energy as a cosmological constant.}
    \label{paramspace}
\end{figure}

The equation of state is constrained to be $-1 \leq w_{Q} \lesssim -0.972$, since we are not considering phantom dark energy models ($w < -1$), and the upper bound is obtained from recent observations of the BAO, the CMB, and SNe \cite{Seo,Brout_2022,rubin2024unionunitycosmology2000,descollaboration2024darkenergysurveycosmology}.
This leads to the constraint
\begin{equation}\label{eq:OmegaKE-BAO-CMB-SNE}
0 \leq \Omega_{\rm KE} \lesssim 9.7\times 10^{-3}.
\end{equation}
Combining Eqs.~(\ref{eq:Q0PhonFinal}), (\ref{eq:average delta E/E}), and (\ref{eq:OmegaKE}), while also imposing the constraint from Eq.~(\ref{eq:OmegaKE-BAO-CMB-SNE}), we obtain the
allowed parameter space for $\Omega_{\rm KE}$ for values of $\alpha$ and $p$ using the liquid drop model shown in Fig. \ref{paramspace}.

\section{Summary}

A novel framework for relating deviations in astrophysical gamma-ray spectra to a pseudoscalar quintessence field $Q(t)$ has been developed. The existence of this pseudoscalar ultralight axion-like quintessence field is posited because of the observational anomalies and theoretical problems that arise with dark energy in the form of a cosmological constant. This work is based on the variations of the pion and quark masses that would result from the coupling of the pseudoscalar field to nucleons, which fits into the study of the variation of fundamental constants more generally. This framework was developed to determine the validity of pseudoscalar models of dark energy specifically. This provides an opportunity to reduce the number of dark energy candidates, of which pseudoscalars make up a significant class.

The main results of this analysis are collected in Table \ref{tab:Results}. The SN 1991T data indicates an energy deviation of approximately $-(0.6\pm0.8)$\%, which is the averaged shift of the first and second excited states of $^{56}\text{Fe}$. This value and its uncertainty suggest that the detected energies are consistent with their terrestrial values. The implied pion mass variation is ($\delta \dot{m}_{\pi}/m_{\pi}) \leq -15\times10^{-11}\text{ yr}^{-1}$. The resulting value of $\dot{Q}_{0}\leq 7\times10^7$ GeV/yr, is at least 11 orders of magnitude smaller than the constant term of the field $Q_{\Lambda}\sim10^{18}$ GeV.

\acknowledgments
We wish to thank Ephraim Fischbach for useful discussions, encouragement, and support for this work.

\begin{table}
    \caption{Deviations and upper limit as indicated by the detected gamma rays of SN 1991T. The uncertainties here are a combination of the statistical uncertainty and the systematic uncertainty.}
    \label{tab:Results}
    \begin{ruledtabular}
    \begin{tabular}{cc}
        Quantity & Result \\
        \hline 
        $\delta E/E$ & $-0.006\pm0.008$ \\ 
        $\delta \dot{E}/E$ & $(1\pm2)\times10^{-10}\text{ yr}^{-1}$ \\
        $\delta \dot{m}_{\pi}/m_{\pi}$ & $\geq -15\times10^{-11}\text{ yr}^{-1}$ \\ 
        $\dot{Q}_{0,max}$ & $\leq 7\times10^7$ GeV/yr \\
    \end{tabular}
    \end{ruledtabular}
\end{table}

\appendix

\section{Energy Deviation for the Rigid Rotor}\label{app:RigDeriv}

In this appendix, we fill in the steps of the derivation of the energy deviation of nuclear energy levels caused by the time evolution of pseudoscalar quintessence, assuming a rigid rotor model. In this case, the energy of an excited state of the nucleus with angular momentum $J$ is
\begin{equation}
    E = \frac{\hbar^2J(J+1)}{2\mathcal{I}}.
\end{equation}
Inserting the moment of inertia $\mathcal{I}$ of an ellipsoid yields
\begin{equation}
    E = \frac{5\hbar^2 J(J+1)}{4A^{5/3}m_p(1+0.31\beta)}\frac{1}{r_0^2}.   
\end{equation}
The variation of the energy with respect a small change $\delta r_{0}$ in $r_0$ is
\begin{equation}
    \delta E = -\frac{5\hbar^2 J(J+1)}{2A^{5/3}m_p(1+0.31\beta)}\frac{1}{r_0^3}\delta r_0.
\end{equation}
The fractional change in energy is then given by
\begin{equation}\label{eq:EDeviationRigInr0}
    \frac{\delta E}{E} = -2\frac{\delta r_0}{r_0}. 
\end{equation}
Using Eqs.~(\ref{eq:PionMassVariation}) and (\ref{eq:RadiusPion}), the fractional energy deviation can be expressed in terms of the QCD angle  $\theta$:
\begin{equation}
    \frac{\delta E}{E} = -2.4 \frac{\delta m_{\pi}}{m_{\pi}} = 0.12\theta^2.
\end{equation}
Assuming the value of the QCD vacuum angle is entirely due to the ALP dark energy field, using Eq. (\ref{eq:QCDandQ}) and then expanding the field $Q$ using Eq. (\ref{eq:TaylorExp}), produces
\begin{equation}\label{eq:dE/ERig}
    \frac{\delta E}{E} \approx \frac{0.12}{f_Q^2}\left(Q_{\Lambda}^2 + 2Q_{\Lambda}\dot{Q}_0\Delta t\right),
\end{equation}
to first order in $\dot{Q}$. However, the first term is constant and renormalizes all energy levels within the nucleus; it does not produce any observable energy deviations as time evolves. To find the time variation of the fractional energy shift, we differentiate Eq.~(\ref{eq:dE/ERig}) with respect to time, obtaining
\begin{equation}
    \frac{1}{E}\frac{dE}{dt} \approx  \frac{0.24}{f_Q^2}Q_{\Lambda}\dot{Q}_0,
\end{equation}
where we have assumed assumed $\dot{Q}_{0}^2\simeq 0$.  Setting $dt \approx \Delta t \equiv t - t_{0}$

\begin{equation}\label{eq:RigRotLim}
   \frac{\dot{E}}{E} \equiv \frac{1}{\Delta t}\frac{\delta E}{E}=\frac{0.24}{f_Q^2}Q_{\Lambda}\dot{Q}_0.
\end{equation}

\section{Energy Deviation for the Nuclear Liquid Drop}\label{app:PhonDeriv}

In this appendix, we derive the time rate of change of the deviation of nuclear energy levels assuming a liquid drop model.  In this case, the energy of the nucleus is associated with quadrupole ($\lambda = 2$) surface vibrations is represented by a number of phonons ($N$) as 
\begin{equation}
    E_N = N\hbar\omega_2,
\end{equation}
where the phonon frequency is determined to be
\begin{equation}
    \omega_2 = \varsigma\sqrt{\frac{1}{r_0^2}-\frac{3}{10}\frac{Z^2e^2}{\varepsilon A^{1/3}}\frac{1}{r_0^3}},
\end{equation}
after inserting the coefficients given in Eq.~(\ref{eq:PhononCoefficients}). In the above equation, $\varepsilon \equiv 4\pi r_0^2S$, which is approximately 18.56 MeV \cite{Feshbach} and $\varsigma$ is defined as
\begin{equation}
    \varsigma \equiv \sqrt{\frac{8\varepsilon}{3mA}}.
\end{equation}
Taking the variation of $\omega_2$ with respect to $r_0$ yields
\begin{equation}
    \delta\omega_2 = -\varsigma \frac{\delta r_0}{r_0} \left(\frac{\frac{1}{r_0^2}-\frac{3}{2}\frac{\xi}{r_0^2}}{\sqrt{\frac{1}{r_0^2}-\frac{\xi}{r_0^2}}}\right),
\end{equation}
where 
\begin{equation}
    \xi \equiv \frac{3}{10} \frac{Z^2e^2}{\varepsilon A}\frac{1}{r_0}.
\end{equation}
Dividing both sides by $\omega_2$ results in
\begin{equation}\label{eq:EDeviationPhonInr0}
    \frac{\delta E}{E} = \frac{\delta\omega_2}{\omega_2} = -\frac{(1-3/2\xi)}{(1-\xi)}\frac{\delta r_0}{r_0}.
\end{equation}
This has the same form as Eq.~(\ref{eq:EDeviationRigInr0}) in Appendix~\ref{app:RigDeriv} except
\begin{equation}
-2\frac{\delta r_0}{r_0} \rightarrow -\frac{(1-3/2\xi)}{(1-\xi)}\frac{\delta r_0}{r_0}.
\end{equation}
Following the same steps that follow Eq.~(\ref{eq:EDeviationRigInr0}), we obtain
\begin{equation}
    \frac{\delta \dot{E}}{E} \equiv \frac{1}{\Delta t}\frac{\delta E}{E}= 0.12\frac{(1-3\xi/2)}{(1-\xi)}\frac{Q_\Lambda\dot{Q}_{0}}{f_Q^2}.   
\end{equation}
\\


\bibliography{apssamp}

\end{document}